\documentclass[sn-chicago]{sn-jnl}

\usepackage{graphicx}%
\usepackage{subfig}
\usepackage{multirow}%
\usepackage{amsmath,amssymb,amsfonts}%
\usepackage{amsthm}%
\usepackage{mathrsfs}%
\usepackage[title]{appendix}%
\usepackage{xcolor}%
\usepackage{textcomp}%
\usepackage{manyfoot}%
\usepackage{booktabs}%
\usepackage{algorithm}%
\usepackage{algorithmicx}%
\usepackage{algpseudocode}%
\usepackage{listings}%
\usepackage{multirow}
\usepackage{array} 
\usepackage{makecell} 
\usepackage{arydshln} 

\theoremstyle{thmstyleone}%
\theoremstyle{thmstyletwo}%

\theoremstyle{thmstylethree}%

\begin{document}

\title[Article Title]{Leading by the Nodes: A Survey of Film Industry Network Analysis and Datasets}


\author*[1]{\fnm{Aresh} \sur{Dadlani}}\email{adadlani@mtroyal.ca}

\author[2]{\fnm{Vi} \sur{Vo}}\email{vv2@ualberta.ca}

\author[2]{\fnm{Ayushi} \sur{Khemka}}\email{khemka@ualberta.ca}

\author[2]{\fnm{Sophie Talalay} \sur{Harvey}}\email{smharvey@ualberta.ca}

\author[2]{\fnm{Aigul} \sur{Kantoro Kyzy}}\email{kantorok@ualberta.ca}

\author[2]{\fnm{Pete} \sur{Jones}}\email{pjones3@ualberta.ca}

\author[2]{\fnm{Deb} \sur{Verhoeven}}\email{deb.verhoeven@ualberta.ca}

\affil[1]{\orgdiv{Department of Mathematics and Computing}, \orgname{Mount Royal University}, \orgaddress{\city{Calgary}, \state{Alberta}, \country{Canada}}}

\affil[1]{\orgdiv{Faculty of Arts}, \orgname{University of Alberta}, \orgaddress{\city{Edmonton}, \state{Alberta}, \country{Canada}}}




\abstract{This paper presents a comprehensive survey of network analysis research on the film industry, aiming to evaluate its emergence as a field of study and identify potential areas for further research. Many foundational network studies made use of the abundant data from the Internet Movie Database (IMDb) to test network methodologies. This survey focuses more specifically on examining research that employs network analysis to evaluate the film industry itself, revealing the social and business relationships involved in film production, distribution, and consumption. The paper adopts a classification approach based on node type and summarises the key contributions in relation to each. The review provides insights into the structure and interconnectedness of the field, highlighting clusters of debates and shedding light on the areas in need of further theoretical and methodological development. In addition, this survey contributes to understanding film industry network analysis and informs researchers interested in network methods within the film industry and related cultural sectors.}




\maketitle

\section{Introduction}
\label{sec1}
\fontdimen2\font=0.56ex
In this paper, we present a survey of published research applying network analysis methods to the study of the film industry. The film industry has long been a rich source of information for network analysis. Use of the Internet Movie Database (IMDb) is prevalent in foundational network studies that require exemplary real-world applications for network theory (\cite{Barabasi1999, Ahmed2007, Madduri2009, Fatemi2012, Gallos2013}). \cite{DiMaggio2011} has identified the ready availability of film industry data via the IMDb API as a key determinant in a range of defining studies. The memorable Six Degrees of Kevin Bacon game (\cite{Fass1996}) embedded the idea of the film industry as a collaboration network in the popular imagination. For academic studies, it was \cite{Watts1999} that used this same dataset to establish a long line of work on small-world collaboration networks based on film data as a use case (\cite{Amaral2000, Rozenfeld2010, Gallos2013}). Given the critical importance of film industry data to the theoretical development of network science, we were interested to switch perspective to understand the ways in which network science has developed a perspective of the film industry as an inherently collaborative sector. In this article, we present the first systematic survey of this area of research, in order to guide network scientists interested in analysing the film industry on how they might usefully approach this sector given what has already been done.

For this survey we are interested in papers that do more than simply use film industry data to test a network methodology or theory. Our focus is on papers that use network analysis~to provide an evaluation of the film industry itself. The film industry is interwoven with manifold connections, involving filmmakers, actors, production companies, distributors, venues, and cinemagoers, each contributing to a rich tapestry of inter-relationships. This intricately structured ecosystem is a compelling domain for network analysis, which holds the potential to reveal crucial insights into patterns of industry collaboration, influence, and resource allocation. Network analysis of the film industry seems especially relevant for making visible the social and commercial relationships that inform the production, distribution, and consumption of the cinema (\cite{Verhoeven2020}), and could be useful within the industry itself, for optimizing marketing strategies, distribution channels, and ultimately enhancing a film’s box-office performance.

Our overarching goal in surveying this literature is twofold: firstly, to establish what has been undertaken in this area, and secondly, to evaluate how scholarship focussed on film industry networks has evolved as an emerging field of study.  In short, our survey of the literature reveals several broad trends and patterns. Initial film network research for example is primarily focused on individuals, with recent papers adopting an expanded scope including production companies and geographic units. Additionally, we observe a disproportionate emphasis on the US film industry in the foundational literature with this bias only amplified by repeated engagement and cross-citation in subsequent papers. We outline a more detailed account of these observations at the conclusion of this essay.

Our aim in providing these insights is to assist scholars interested in utilizing network methods to explore the film industry, as well as other cultural industries, by demonstrating where existing scholarship has clustered into emergent debates and approaches, and where new research might be directed. In particular, we highlight the research lines that have advanced the least, and identify gaps in the existing literature that future network analyses of the film industry could address.

\section{Scope and Analytical Strategy}
\label{sec2}
\fontdimen2\font=0.56ex
The scope of our survey includes any scholarly work using network analytic methods~to study film, with some defined exclusions. First, to ensure that we were able to comprehensively read and review the papers, we restricted our review to English language publications. Second, we did not consider theses for our review. Third, we excluded papers which were not substantively interested in film but used a film-related dataset purely~to illustrate a methodological development in network analysis. Fourth, we excluded papers using network approaches to represent film narratives as networks of interactions or co-occurrences between characters, as there already exists a comprehensive survey of the character networks literature which covers this type of study (see \cite{Labatut2019}). Finally, we excluded network-oriented papers which do not attempt to represent the networks they describe as data. This helped us keep a check on the many varied ways in which ideas of “networks’’ are used discursively in different disciplines, many of which do not align with the type of network analysis we survey in this review. As a consequence of this final criterion, we do not include papers here such as \cite{Coe2000} which provide rich and valuable qualitative evidence of the importance of personal networks in the economic geography of film production, but do not contain any data-driven network analysis per se.

We used a variety of search systems and keywords (“imdb”, “movie”, “film industry” in combination with “networks”, “SNA”, “network analysis”, “small world”, “centrality”) to identify papers across a range of platforms (Google Scholar, ResearchGate, Academia.edu, university library catalogues and journal aggregator databases). Once we had compiled an initial list of papers returned from keyword searches, we chained the references in these papers to find further papers. Additionally, we examined each author’s publication list to ensure no relevant papers were overlooked. New papers appeared as we were searching, so we set a publication expiry date for the end of the calendar year 2022.

\begin{figure*}[!t]
    \centering
    \subfloat[Number of papers reviewed by node type.]{\includegraphics[width=0.49\textwidth]{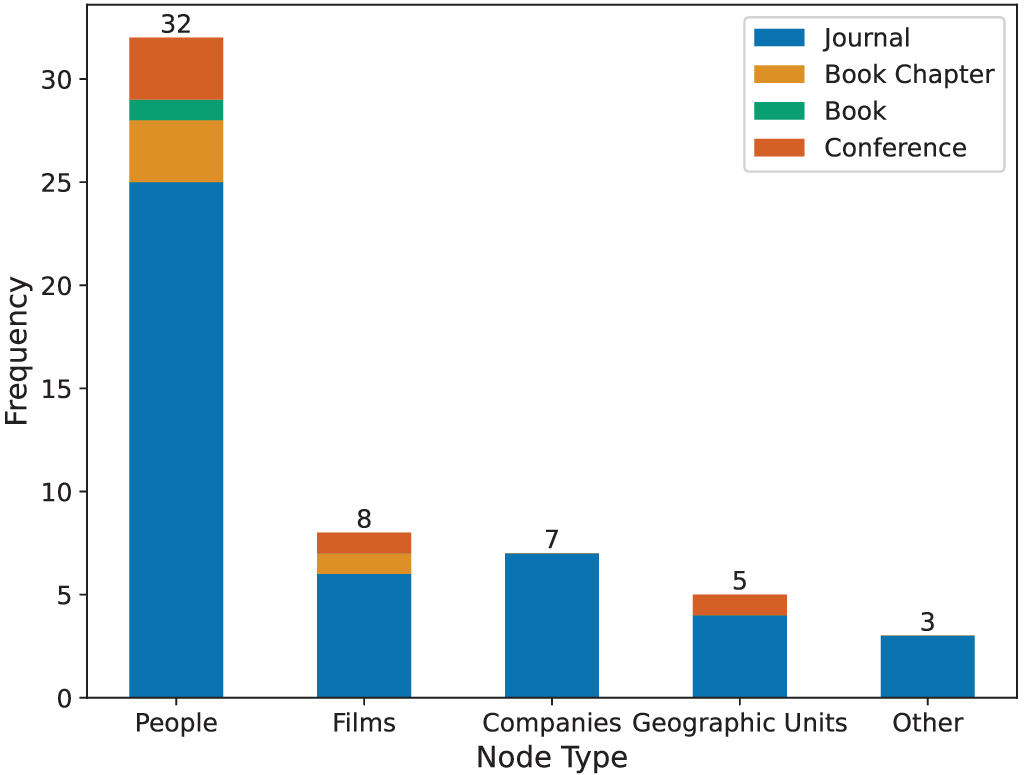}\label{fig1a}}
    ~
    \subfloat[Papers based on node type.]{\includegraphics[width=0.51\textwidth]{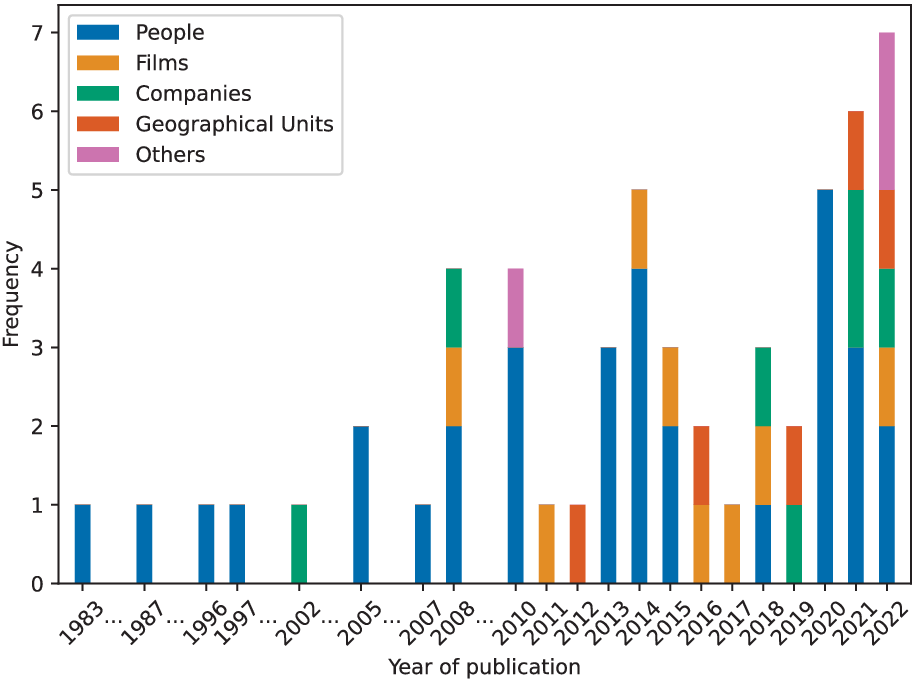}\label{fig1b}}

    \subfloat[Citations based on node type.]{\includegraphics[width=0.51\textwidth]{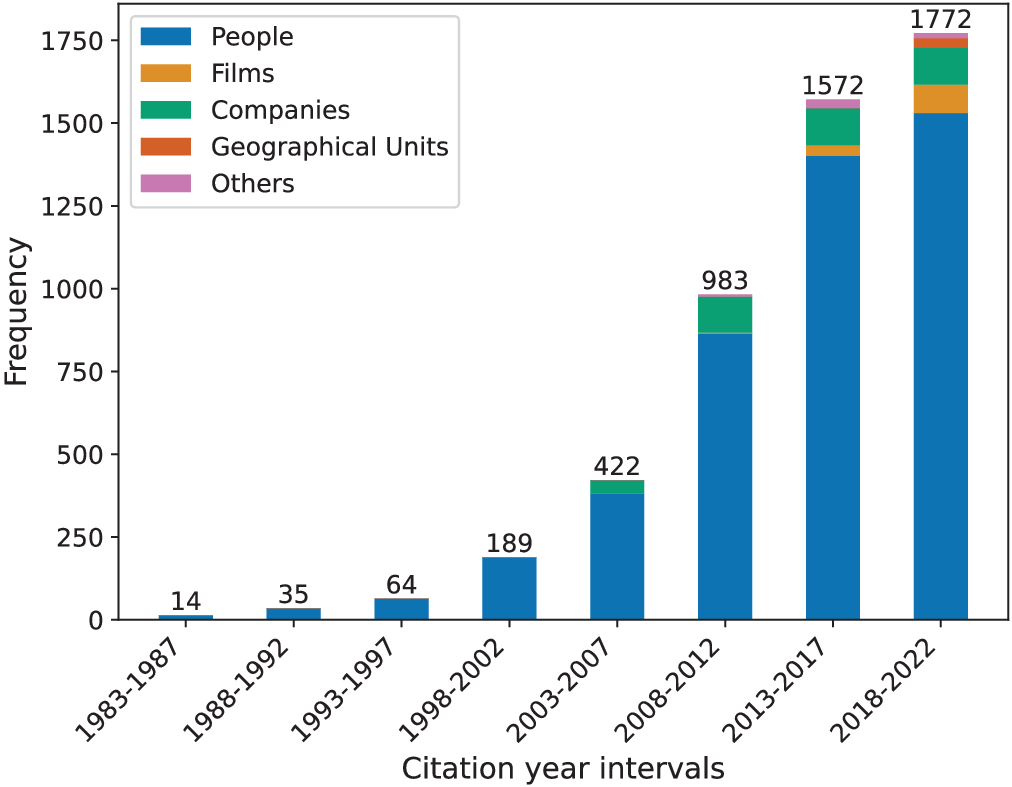}\label{fig1c}}    
    \caption{Frequency distributions of (a) the number of papers reviewed by node type, (b) the papers based on node type from years 1983 to 2022, and (c) the citations per each node category over the years. Note that the numbers by node type do not sum to 51 as papers may analyze more than one type of node.}
    \label{fig1}
\end{figure*}

Our comprehensive search efforts yielded 51 papers for review. During the review process, we documented information about the type(s) of nodes represented in the network and the publication format (e.g. journal articles, book chapters). These are summarized in \figurename{~\ref{fig1a}}. After the search was complete, we recorded the total number of citations of~each paper according to Google Scholar, providing an additional metric for assessing~the impact and significance of the reviewed literature. \figurename{~\ref{fig1b}} showcases the frequency~of publications over time based on the node type. Notably, there is an escalating~interest in network heterogeneity in recent years, underscoring that network research of film industries is ongoing and continuing to expand. The citation distribution in \figurename{~\ref{fig1c}} clearly shows a growing focus on networks of individuals over the examined period, while research concerning networks comprised of geographic units remains in its nascent phase.

\section{Survey of the Literature}
\label{sec3}
\fontdimen2\font=0.56ex
Survey papers typically adhere to the established structure of their respective fields. However, the literature we review spans multiple disciplines, including film industry studies, business studies, cultural policy studies, sociology, and computational sciences. These disciplines and research fields are structured very differently which complicates the task of conducting a coherent survey. One solution would be to categorize the papers by field or discipline. However, discerning the appropriate disciplinary classification for certain papers is not a straightforward process. Another conventional approach would be to distinguish papers based on different network methods. Our emphasis however was on social network papers evaluating the film industry rather than studies simply using film industry data to explicate or illustrate a network method. Excluding purely method-driven papers made this organisational principle less useful.

To address these challenges, we devised a novel solution involving the classification of papers based on node types. Through systematic examination, we found that scholarship exploring networks of individuals involved in the filmmaking process is often substantively and methodologically different from research focused on connections between films, organizations, events, or geographical units. Although a small number of research papers investigate distinct networks with different node types, to our knowledge there has been no significant study to date that considers the interdependencies among multiple types of nodes within film networks. 

\subsection{Node Type: People}
\label{sec3.1}
\fontdimen2\font=0.54ex
\subsubsection{Careers and Networks
}
In a groundbreaking work on the careers of screen composers in Hollywood, \cite{Faulkner1983} demonstrates that the industry is dominated by a select group of highly productive and prominent individuals who hold significant control over resources, career and opportunities. Through the use of statistical analysis and personal interviews with both established and peripheral composers, Faulkner traces a network linking filmmakers (employers) and composers (employees) to reveal patterns of work distribution and labor division. Building on this study, \cite{Faulkner1987} uncover connections between~independent contractors in the film industry based on their co-working and contracting patterns. Using data from 2,430 films produced between 1965 and 1980, they identify trends in the repeated connections among participants with similar levels of cumulative productivity, measured by earnings, Oscars film credits. They note that film producers tend to hire directors with a higher earning history, and lower-earning producers work with lower-earning directors. Similarly, directors with a successful track record of high-grossing films contract actors and cinematographers with similar histories. For Faulkner and Anderson these patterns demonstrate that Hollywood's collaboration network is based on reputation and that career navigation is driven by interpersonal signals and connections.

In another paper on the role of social networks in project-based careers, \cite{Jones1996} addresses the questions: How are careers and work structured in the dynamic and decentralized film industry of Hollywood? What expertise and knowledge is vital for career success when participation is renegotiated for each project? How do careers and projects align to maintain a networked organization? Are boundaryless (project-based) careers primarily based on temporary transactions, or repeated interactions that foster learning and innovation? The author analyzes the project network in the US film industry (1977-79) using $k$-core analysis. They show that to achieve success, individuals aim to acquire new technical skills and/or form connections to become part of more production teams. Similarly, \cite{Jones1997} explore how independent contractors navigate the US film industry and how their position within the core or periphery of the network impacts their careers. The authors also examine labor market stratification and segmentation in the industry between 1977-79. Their study was informed by in-depth interviews with five industry professionals, along with publication records and film credits. Findings highlight fluid entry and exit points for contractors and companies. The authors perform a $k$-core analysis which demonstrates that major studio-affiliated contractors in the central core have better opportunities for high-paying and challenging film projects.

Drawing on this earlier work, \cite{Delmestri2005} explore the importance of reputation and collaboration networks on movie project success. They hypothesize that economic reputation (past box office success) affects commercial achievement, while artistic reputation (favorable critical reviews of past films) influences artistic merit. For network effects, they turn to organizational literature to hypothesize that strong vertical ties (dyadic ties between the director and the distributor and producer) will be positively associated with economic success, while strong horizontal ties (dyadic ties between the director and scriptwriter, director of photography and lead actor) will be negatively associated with artistic merit. Their methodology combines interviews with 14 professionals in the Italian film industry and quantitative data on Italian films from 1990 to 1998. The authors find support for each of their hypotheses, though the negative effect of strong horizontal ties on artistic merit is only marginally significant.

\subsubsection{Networks, Stability and Innovation
}
\cite{Ferriani2005} present the film industry as an example of temporary project firms that maintain stability despite their transitory nature. They study enduring interpersonal relationships within the industry that persist after the completion of individual projects and the disbandment of project firms themselves. Collaborations between various participants in the industry (such as directors, sound mixers, and producers) reveal underlying factors driving organizational knowledge and practice. The paper uses IMDb (cross-checking against the Alan Goble Film Index) to locate and analyze movies distributed in the United States by the seven major studios and the two largest independent subsidiaries (Miramax and New Line) between 1995 and 1999. They find that the repeated collaborations in the industry give rise to latent networks, which counter the lack of~permanent organizational structures. The authors acknowledge they cannot definitively determine whether repeated collaborations occur due to friendship and consider their findings to be suggestive rather than conclusive.

\cite{Perretti2007} analyze film genre innovation in the film industry by assessing the introduction of new genres and their combinations. They hypothesize that a~higher proportion of newcomers and a higher proportion of new combinations of both newcomers and old-timers would be positively associated with innovation. The authors tested these hypotheses on the U.S. movie industry by analyzing 6,446 titles produced by major Hollywood studios from 1929 to 1958. The method involved classifying films as single-genre or multi-genre and classifying production teams as combinations of newcomers, newcomers and old-timers, or old-timers. They concluded that the rate of innovation~was indeed increased by the introduction of newcomers, and that innovation was further increased when teams combined newcomers with old-timers. The study also found that executive level individuals (managers) had little impact on innovation. 

In a particularly influential paper, \cite{Cattani2008} introduce a new perspective on individual-level creativity in which they propose that individuals positioned between the core and periphery of their social network are more likely to achieve creative outcomes. This perspective can also apply to teams, where individuals in a team that have members from both ends of the core/periphery spectrum have an advantage. The study focuses on Hollywood between 1992 and 2003 and uses data of all core crew members who worked in movies distributed by the eight major studios. The first hypothesis is that individual creative performance has an inverse U-shaped relationship with their coreness, implying that intermediate positions yield the highest creative performance. The second hypothesis is that the relationship between individual creative performance and team coreness is also an inverted U-shape, with optimal creative performance achieved by individuals in teams with a moderate level of coreness. The study concludes that individuals bridging core and periphery boundaries enjoy enhanced creative performance, and moderate levels of coreness are beneficial. Teams with a balanced core-periphery mix experience similar benefits. The authors underscore the industry-specific context and acknowledge that creativity is outcome rather than process-focused in this study.

\subsubsection{Networks and Success Prediction
}
\cite{Krauss2008} also focus on understanding film performance outcomes in their investigation of whether opinions expressed by IMDb forum users can be used to predict Academy Award nominations and box office success of unreleased films. The first model used the “Oscar Buz” subforum of IMDb and determined the likelihood of a film receiving an Academy Award nomination by computing three factors: the Intensity Index, the Positivity Index, and the Time Noise Factor. The second model analyzed 20 unreleased films and created a “Buzz Mode” from the Intensity, Positivity, and Trendsetter Indexes. The Trendsetter Index was calculated by identifying subforum users with the highest betweenness centrality in the “Previews and Review” subforum and normalizing their number on a scale of 0 to 1. Results affirm both hypotheses:  positive “Oscar Buzz”~discussions are related to Oscar nominations, and positive discussion on the “Previews and Reviews” subforum correlated with higher box office revenue. 

Box-office prediction is also the focus of a study by \cite{Karpov2022} which aims to predict a movie’s success by utilizing its position within the community structure of the international film industries alongside conventional film-level variables such as language, country, rating and genre. Using IMDb data, the authors construct a network connecting actors, casting directors, talent agents, and directors based on co-affiliations with movies. They demonstrate that incorporating additional information derived from this network can improve the accuracy of predicting a movie’s success by up to $6\%$, depending on the classification method. They also find that the importance of a casting director is greater than that of actors in terms of predicting a movie’s success, which offers the opportunity to predict a movie's success at an early stage, even before actors are approved for roles.

An earlier analysis by \cite{Meiseberg2013} retrospectively examined the relationship between a film’s creative team composition and its success in domestic and international markets. They develop hypotheses about team diversity in terms of personal attributes, film locations, and social ties, and test them using ordinary least squares (OLS) regression models. They found that diverse nationalities and star status, as well as multiple film locations, are the strongest predictors of export success. For domestic success, positive factors include diversity in the age and tenure of team members, and the number of connections in the people-film network. Gender diversity however, was negatively associated with domestic success. They conclude that the right mix of team attributes, experience, and backgrounds influences success, and reducing cultural specificity enhances international market success.

\cite{Viana2014} also study film success in relation to the historical development~of team composition in the film industry by examining topological (small world coefficient, betweenness, closeness, local clustering coefficient) and non-topological (previous experience) properties of directors, producers, and writers. Using IMDb they identified a set of 3,006 films from the late 1800s to 2013 that had received a minimum of 25,000 user votes and including films from the IMDb top 250 list. They then analyze the correlation between these properties and film success, measured by gross income and ratings. They discover that non-topological properties have a stronger relationship with film success, while topological properties have a weaker correlation. Additionally, they find that teams with less experienced members tend to perform better, indicating that bringing in new talent can improve a team’s success. 

Rather than define success in terms of box-office income, \cite{Rossman2010} use data on Academy Award acting nominations to examine the relationship between artistic achievement and collaboration in the film industry. The authors analyzed actors in films released between 1936 and 2005, while controlling for factors such as the actor's personal history and basic film traits. Data from the Academy of Motion Pictures Arts and Sciences (AMPAS) and the Internet Movie Database (IMDb) was used to create a dataset of~the top 10 credited actors. Only films eligible for Academy Awards from 1936 to 2005 were analyzed, a total of 16,392 films featuring 147,908 performances by 37,183 actors. Of these, only 1,326 performances received Oscar nominations and 279 wins, which is less than $1\%$ of the total performances. Using statistical analysis and logistic regression, the study found that actor's status, as measured by their asymmetric centrality in the network of screen credits, is a strong predictor of their chances of receiving an Oscar~nomination. Additionally, actors are more likely to receive an award when they work with elite collaborators. 

\cite{Ebbers2010} aim to understand the impact of performance-based~reputation on the development of alliance networks in the Dutch film industry over time. The authors use stochastic actor-oriented models to examine the impact of an actor’s reputation, which is based on their previous performance, on the formation of alliances, while controlling for constant actor attributes and network position. They differentiate between individual reputation and composite reputation, considering past organizational involvement and partner reputations. Results show that an actor's reputation (with a focus solely on artistic reputation derived from reviews) and their proximity in past alliance networks are strong indicators of alliance formation. The study suggests that actors with similar reputations are more likely to form alliances with each other. The authors performed additional analysis of the impact of commercial reputation, based on a return-on-investment measure using box-office and budget data, but no significant effects were found.

\subsubsection{Historical Studies
}
Two studies use data related to the Hollywood “blacklist” era in which alleged communist sympathizers were investigated by the House Un-American Activity Committee (HUAC). \cite{Pontikes2010} study the role of network diffusion processes for stigmatization. They propose hypotheses about the effect of associating with a later-blacklisted artist on work opportunities. Testing with logit regression models, they found that mere association with a blacklisted artist reduces the odds of working in the next year, and that only one instance of mere association was enough for this negative effect to apply.~The effect is stronger for actors collaborating with blacklisted writers, suggesting stigma spreads through diverse connections. They also observed that high-profile actors are not immune to the effects of stigma. In a related study, \cite{Negro_2015} explore whether niche overlap (two social actors in the Hollywood labour market with similar professional and demographic features) is connected to the act of naming somebody in the HUAC trials. Their hypothesis is that niche overlap is positively related to the likelihood of discrediting. They use data from published reports of the public hearings to identify potential discreditors and their actions in a binary discrediting network. The set of people is limited to actors, directors, writers, and producers involved in commercial feature-length films up to the conclusion of the hearings. They use a naming network, a collaboration network, and combinations of these to define relational patterns associated with the act of discrediting. Employing a multi-level logit model with crossed random effects, and while controlling for various factors including prior collaborations, they found compelling evidence linking niche overlap to an increased likelihood of discrediting.

\subsubsection{Gatekeeping}
Network analysis has proved especially valuable to several authors exploring the importance of gatekeeping as both a commentative and as an employment practice in the film industry. \cite{Cattani2013} test the hypothesis that industry peers who act~as evaluators tend to reinforce dominant social beliefs and norms, and are more likely to acknowledge core members over peripheral network members. Using Hollywood data from 1992 to 2004 they constructed film collaboration networks and analyzed the bipartite affiliation networks by creating an individual-by-movie matrix. The authors also~created an “accumulative relational profile” by using moving windows of three-year intervals. Their findings suggest that social relationships play a crucial role in determining how talent is perceived and acknowledged. \cite{Cattani2014} extend this study by drawing on field theory to hypothesize that peers in the film industry favor creators at the core of the collaboration network, while film critics may prefer those at the periphery. These hypotheses are tested using a dataset of over 12,000 crew and cast members who worked on movies distributed by eight major studios in the United States from 1992 to 2004. The dataset includes information on directors, writers, leading and supporting actors/actresses, editors, cinematographers, and production designers, sourced from IMDb. They use a discrete choice model to analyze award decisions by organizations and estimate the choice set of eligible professionals. Collaboration networks are created based on person-film affiliations and analyzed using a core-periphery algorithm. Node classification as~core or periphery is then used as a predictor in the models. Results show that being in the~periphery negatively affects peer recognition but has a slight positive association with recognition by critics, controlling for other factors. 

The impact of network closure on gender inequality in a project-based labor market was examined in an important study by \cite{Lutter2015} of Hollywood film productions between 1929 and 2010. Lutter analyzes data on 1,072,067 film performances by 97,657 actors, $30\%$ of which are women, taken from IMDb. The author utilizes Cox regression models, a type of survival analysis, to estimate the risk of career failure, represented as a binary variable in which a value of 1 indicates an actor’s final appearance. The results find that gender equality is more attainable in a network structure that is open and diverse, leading to statistically insignificant differences between men and women in terms of career failure when there is no team cohesion present. \cite{Verhoeven_Kaska2020} build on Lutter’s research on gender inequality in the film industry and analyze the role of social networks in perpetuating this inequality. The study uses data from the Australia, German and Swedish film industries and confirms previous findings that women tend to remain on the periphery of networks unless they can create ties with key players. The authors further propose control mechanisms to address this, such as establishing employment connections between women and men who are key players, but acknowledge that these interventions may not always be desirable in practice. The authors argue that aggregated statistics on gender inequality are not enough to propose meaningful change and that research should focus on understanding gender inequality in terms of uneven social relations.

Advancing the analysis of gendered social relations in the film industry, \cite{Fanchi2020} provide an analysis of the impact of digital transformation on the Italian cinema industry, particularly on the position of women in production structures. They find that women have a peripheral role in the Italian cinema industry and are less likely to work on films with higher budgets and prestige. They note that women are numerically underrepresented, but that the problem of women creatives in the industry is not just about number but also power and position; women are not only excluded, those that are able to enter the industry occupy less central positions. This imbalance can have cultural, symbolic, and economic repercussions on industry competitiveness and renewal. Overall, the study suggests that digital transformation has not been sufficient to eradicate the gendered logics of Italian cinema, and bolder interventions are necessary to promote a more inclusive production system. The paper proposes that policy reforms need to place more emphasis on connecting the clusters of women creatives with the larger and more powerful regions of the Italian film industry creative network.

\cite{Juhasz2020} combine gatekeeping analysis with a focus on creative success to examine how core creators benefit from brokerage in core/periphery networks~using a unique, open access dataset of Hungarian feature films. They find that creators who bridge the core and periphery in the collaboration network have the highest likelihood of creative success. This dual core-broker position provides complementary benefits~to central creators. The study challenges the core/periphery trade-off using a measure called gatekeeping to identify brokers between the core and the periphery~of the network. The~analysis is based on the three-way interaction effects of coreness,~edge betweenness-based brokerage, and gatekeeping to understand their joint impact on creative success. 

\subsubsection{Specific Industry Studies
}
Two papers analyse the adult film industry as a specific case study. \cite{Gallos2013} utilize the movie co-appearance network from IMDB, in which actors are connected if they have appeared in the same film, to demonstrate how to uncover fractality in a small-world network. They analyzed a subset of the IMDB network containing “Adult” films, arguing that this subset demonstrates a clear transition from small-world to fractal and modular structures because the genre is more recent, more focused in time, and isolated from other genres. The study uses the box-covering technique and renormalization to conduct the analysis. Results show that the strength of links, or the number of times actors collaborate, can significantly impact the network topology. When all actors are linked, the network becomes small-world and lacks a proper modular structure. However, when a threshold is imposed on the minimum number of links between two actors, the network becomes naturally fractal. \cite{Lüdering2018} also examines adult film industry networks to evaluate the link between collaboration network position and career duration in the adult film industry. Using data from the Internet Adult Film Database, Lüdering creates yearly networks of adult film performers based on co-appearance in films and then, drawing from social capital theory, applies eigenvector centrality to analyze how network positioning impacts career length. Survival models include centrality, connection to the main network component, age at entry, and gender as explanatory variables. Different approaches to baseline hazard distribution are considered, including stratified/unstratified Cox regressions and parametric models. In all cases, centrality lowers exit risk, while disconnection from the main network raises it.

There are a series of studies that delimit their exploration to the networks found within national cinemas. Two interrelated studies use social network analysis to study the role of key individuals in the Afrikaans film industry. \cite{Senekal_2014_1} analyse~the structural importance of filmmaker Jamie Uys in the industry, while \cite{Senekal_2014_2} focuses on director Pierre de Wet. Both papers use the same data set on the Afrikaans film industry and deploy very similar analytical methods. The dataset for the Afrikaans film industry includes entities such as films, directors, film editors, producers, writers, cinematographers, music composers, production companies, and distribution companies from 1916 to 2013. The initial information was obtained from a 1982 study by Le Roux and Fourie, and additional details were then gathered from IMDb and production company web archives. The different entity types form a multimodal network, and all analyses are based on different subsets of this network. Senekal finds that Pierre de Wet had a high level of centrality and was able to spread knowledge throughout the industry network. Senekal argues that the high degree centrality of de Wet and many of his collaborators had a significant impact on the flow of information in the film industry as it allowed for the accumulation of experience through working on a large number of films, as well as the transfer of experience through contact with many other individuals who also have many connections. Similarly, Senekal and Stemmet find that both Uys and his films have high betweenness centrality, making him a hub in the industry.

Two papers in our survey focus specifically on the Turkish film industry. \cite{Abanoz2020} uses data from IMDb to explore the small world properties of the actor co-occurrence network in Turkish cinema. The study uses closeness and average path length to identify the actors who are at the centre of the small world structure in the one-mode actor-actor projection of the actor-film affiliation network. They find that Süheyl Eğriboz is the most central by these measures and argue that he was the Turkish equivalent of Kevin Bacon in that everyone in Turkish cinema is either directly connected with him or indirectly connected to him through few intermediary collaborators. In another paper examining Turkish film industry networks, \cite{Savk2021} critique the traditional approach to film historiography that focuses on films and major players. Using the Yeşilçam subindustry in the 1960 and 70s as a case study, the authors employ network analysis to shift the focus onto the connections among writers, directors and producers based on patterns of collaborating on films. They identify two distinct clusters in the resulting network visualizations, which they argue accurately reflect the split between mainstream and lower-budget modes of production in that era.

\cite{Neuberger2020} explores a network depiction of the Soviet film industry from 1918 to 1953. Using data from the Russian film database kino-teatr.ru, the author scrutinizes director-actor connections primarily using eigenvector centrality and community detection. The research reveals that the director-actor network starts with divisions along ethnic and geographic lines but gradually integrates over time. Furthermore, the author notes that those with high eigenvector centrality are not the typical figures identified as significant by Soviet cinema scholars and argues that a network approach provides a distinct viewpoint on the composition and development of the Soviet film industry.	 

Finally, \cite{Noroozian2022} analyze an original dataset of Iranian actors who participated in the Fajr Film Festival from 1998 to 2020 to investigate yearly collaboration networks among the actors. Their analytical focus is on the decreasing diversity of Iranian film genres over time and the impact of persistent collaboration between key actors on audience preferences, stability of genres, and box office measures. Findings suggest that network density negatively affects box office, but one particular type of film, the social problem genre, has stabilized due to the continuous cooperation between core actors. Furthermore, there is a positive correlation between the importance of actors in the network and their popularity on Instagram, indicating that people follow those central in the actors' network. Table~\ref{table_1} highlights some of the main features of the research efforts reported in this sub-section.

\begin{table}[!t]
\caption{Feature summary of reported film network analysis with people as nodes.}
\begin{tabular}{@{}lp{1.45cm}p{2.33cm}p{4.4cm}@{}}
\toprule
Reference & Team Size\footnotemark[1] & Data Size\footnotemark[2] & Evaluation Methodology \\
\midrule
\cite{Faulkner1987}  & 2 (M)  & $2,430$ films  & - Regression \\
\cite{Jones1996} & 1 (W) & $606$ films  & - Component, $k$-core analysis\\
\cite{Jones1997} & 2 (W) & $606$ films   & - Component, $k$-core analysis \\
\cite{Delmestri2005} & 3 (M) & $705$ films   & - Regression \\
\cite{Ferriani2005} & 3 (M) & $762$ feature films &  - Blau’s index, t-test \\
\cite{Perretti2007}  & 2 (M) & $6,446$ feature films &  - Regression \\
\cite{Cattani2008} & 2 (M) & $2,137$ films  &  - Regression \\
\cite{Krauss2008} & 5 (M) & - &  - Correlation analysis \\
\cite{Rossman2010} & 3 (M) & $16,392$ films &  - Regression \\
\cite{Pontikes2010} & 3 (W) & $5,712$ feature films &  - Weighted logit regression \\
\cite{Ebbers2010} & 2 (M) & $233$ films &  - Correlation analysis \\
\cite{Meiseberg2013} & 2 (W) & $180$ films &  - Regression  \\
\cite{Gallos2013} & 4 (M) & $39,397$ adult films &  - Renormalization group analysis \\
\cite{Cattani2013} & 2 (M) & $2,297$ films &  - Logit regression  \\
\cite{Cattani2014} & 3 (M) & $2,297$ films &  - Logit regression  \\
\cite{Senekal_2014_1} & 2 (M) & - &  - Centrality analysis \\
\cite{Senekal_2014_2} & 1 (M) & - &  - Centrality analysis \\
\cite{Viana2014}  & 4 (M) & $3,006$ films &  - Network analysis \\
\cite{Lutter2015} & 1 (M) & $369,099$ films &  - Regression  \\
\cite{Negro_2015} & 2 (M) & - &  - Multi-level logit regression  \\
\cite{Lüdering2018} & 1 (-) & $102,871$ adult films &  - Survival analysis, regression \\
\cite{Juhasz2020}  & 3 (M) & $7,672$ creators &  - Regression, centrality analysis \\
\cite{Fanchi2020}  & 2 (W) & $3,542$ films &  - Centrality analysis \\
\cite{Verhoeven_Kaska2020} & 7 (W) & $4,054$ creators &  - Network analysis \\
\cite{Abanoz2020} & 1 (M) & $9,441$ films &  - Network analysis \\
\cite{Neuberger2020} & 1 (W) & $4,810$ directors/actors &  - Network analysis \\
\cite{Savk2021} & 2 (M) & $5,595$ films &  - Network analysis \\
\cite{Karpov2022} & 2 (M)   & $85,855$ films (Y) & - Network and predictive analysis  \\
\cite{Noroozian2022} & 3 (M)   & - & - Network analysis  \\
\botrule
\end{tabular}
\footnotetext[1]{Refers to the number of authors, followed by the gender (M/W) of the lead author.}
\footnotetext[2]{Refers to the target network and whether the dataset used is publicly available (Y).}
\label{table_1}
\end{table}

\subsection{Node Type: Films}
\label{sec3.2}
\fontdimen2\font=0.54ex
\subsubsection{Critical Success Indicators
}
Less prevalent than people nodes but still significant, is the use of film titles as network nodes. Much of this research focuses on measuring the success and/or impact of particular films or collections (e.g. genres) of films. \cite{Meiseberg2008} provide an early example of this approach using data on the German film industry. Their dataset is drawn from the 10 highest-admission German films each year as recorded by the German national film funding agency, amounting to 111 films released between 1993 and 2004. From these data they first construct two-mode actor-movie networks, which they project down to a one-mode movie-movie network in which ties represent shared actors. They conduct regression analyses to test hypotheses about the effects of various film-level factors on the performance of German films, measured in terms of admissions. They~include variables in this model for the degree of movies and the local clustering coefficient in the movie-movie network. They find that degree - i.e. the number of movies with which the focal movie~shares actors - is positively associated with movie success, while there is no significant effect for the local clustering coefficient.

Seeking to understand the context for different kinds of film success, \cite{Miller2011} compares the funding models and network patterns of commercially successful and critically acclaimed US films. The study scrutinizes co-production network structures and how they evolve with production changes focusing on high-quality film funding and network structures in domestic and international settings. The role of inter- and intra-nationality in company relationships and their impact on film success is explored. The paper’s findings show that the network structures of both commercially successful films and critically acclaimed ones are significantly similar, being less centralized, clustered and dense than anticipated. However, critically acclaimed films exhibit greater collaboration and network “penetrability” compared to their high-grossing counterparts.

\subsubsection{Interconnected Movies
}
A notable group of film-oriented network papers explore the impact, influence and interconnection of films with another through “on-screen” indicators of intertextuality. These papers primarily draw on the user-submitted “Connections” field on a film’s IMDb page, which allows users to establish six types of connections among movies, including features, references, follows, spoofs, remakes, and spin-offs. For example, \cite{Spitz_2014} use this information to construct a network of film “citations” to evaluate cinematic impacts and milestones. Films are represented as nodes and citations as edges. Centrality indices, such as weighted out-degree, temporal degree, influence time, subtree, start, and propagation, are used to measure impacts. The films are then ranked based on their position in the network and compared to various “greatest film” lists.~The authors find that films with long-term impact, as measured by film citations, do not~always correspond to high ratings or awards. 

\cite{Wasserman_2014} explore the correlation between IMDb user scores and film budgets and gross earnings. The authors construct a film network based on IMDb’s “Connections” section for each film. During network construction, they notice a significant bias toward U.S. films in terms of incoming and outgoing connections. To mitigate this bias, they narrow their dataset to exclusively include U.S. films thereby avoiding any confounding factors associated with the country of production. Their findings reveal a strong correlation between the number of votes and economic statistics, particularly the film’s budget. They conclude that the quantity of ratings is a reliable indicator of a film’s prominence.

Also seeking to explain film success, \cite{Bioglio_2017} introduce the “influence score” as an innovative metric for evaluating a movie’s performance. This method differs from traditional approaches reliant on economic sales data or subjective critic~assessments, which can be influenced by marketing strategies, inflation, diverse distribution platforms, and personal biases. The influence score hinges on a film’s ability to inspire subsequent works through creativity, innovation, and franchise-building. The study investigates intricate patterns and trends related to a film’s release year, genre, and production country. Again, the authors define a citation network where nodes represent movies, connected by directed edges whenever a connection (per IMDb) exists. The study employs centrality scores, including in-degree, closeness, harmonic, and PageRank measures, implemented using Python and the NetworkX library, to evaluate the network’s dynamics. The findings indicate that the top 20 influential movies, based on influence centrality, predominantly originate from the pre-1940 era, highlighting the enduring impact of classic films. Many of these classics come from Anglo-Saxon countries, emphasizing the influence of American, French, Italian, and German cinemas. In a related publication, \cite{Bioglio_2018} argue that citations between films offer a more dependable measure of a film’s impact than assessments by film critics, economic data, or historical comparisons. 

\cite{Shin_2022} develop a variation of “citation” measures by examining the evolution of movie genres and subgenres and their influence on popular culture through the analysis of memes as indicators. They introduce the “meme score”, determined by machine learning algorithms analyzing user reviews and ratings for both movies and tags. Each tag is treated as a meme, comprising elements of a film. Networks of movies are then created by connecting films with similarities that exceed a certain threshold. The study reveals that cross-genre connections grow over time, with successful films spawning new memes. However, the methodology lacks clarity regarding the precise generation of these networks, aside from the use of ML algorithms.

Film industries are notoriously risk averse and the ability to predict box office is a longstanding research problem. \cite{Yahav_2016} emphasizes the importance of understanding consumer choice in predicting movie demand, asserting that movies with similar appeal should exhibit similar demand patterns. To enhance box office success forecasting, they introduce an automated technique that leverages a similarity network. This~technique measures several aspects of demand structures, including decay rate, time of the first demand peak, per-screen gross value at peak time, existence of a second demand wave, and time on screens. Their argument highlights the superior predictive power and robustness of models incorporating a similarity network compared to those that do not. Table~\ref{table_2} summarizes key features of the research endeavors discussed in this subsection.

\begin{table}[t]
\caption{Feature summary of reported film network analysis with films as nodes.}\label{tab1}%
\begin{tabular}{@{}lp{1.45cm}p{2.33cm}p{3.9cm}@{}}
\toprule
Reference & Team Size\footnotemark[1] & Data Size\footnotemark[2] & Evaluation Methodology \\
\midrule
\cite{Meiseberg2008} & 2 (W)   & $111$ films  & - Network analysis, OLS regression   \\
\cite{Miller2011} & 1 (W)   & $110$ production companies  & - Network and correlation analysis   \\
\cite{Wasserman_2014} & 6 (M) & $32,636$ films   & - Correlation analysis \\
\cite{Spitz_2014} & 2 (M) & $40,000$ films &  - Centrality analysis \\
\cite{Yahav_2016}  & 1 (W) & $2,240$ films &  - Regression \\
\cite{Bioglio_2017}  & 2 (M)  & $65,000$ films  & - Centrality analysis \\
\cite{Bioglio_2018} & 2 (M) & $47,000$ films  & - Centrality analysis \\
\cite{Shin_2022} & 3 (M) & $10,380$ films  & - Machine learning \\
\botrule
\end{tabular}
\label{table_2}
\end{table}

\subsection{Node Type: Companies, Geographical Units, and Others}
\label{sec3.3}
\fontdimen2\font=0.54ex
In this subsection, we compile research papers whose network nodes represent companies, geographical units, or other relevant entities in the film industry. These papers are summarized in Table~\ref{table_3}.

\subsubsection{Networks and Box-Office Prediction 
}
The interest in the value of networks for predicting film industry performance~continues in this section with \cite{Doshi_2010} who explore the effectiveness of social~network analysis and sentiment analysis within the context of predicting Hollywood Stock Exchange (HSX) pricing trends during the first four weeks following a movie release. This prediction process operates by predicting the potential financial success of the film, predicting the audience’s reception and likability of the film, and estimating the film’s actual earnings within the first month after its release. To facilitate these predictions,~the authors harness web metrics extracted from IMDb, Rotten Tomatoes, movie quotes~from HSX, and box office collection data from Box Office Mojo. They conduct sentiment analysis on forum posts within IMDb to discern positive and negative connotations and assess the film’s popularity. Additionally, they gather data on movie titles and monitor their relevance within the blog space, termed the “information sphere”. Utilizing this~data, they calculate betweenness centrality to quantify the buzz and attention a movie garners on the Internet. To refine their predictions, a combination of multilinear and non-linear regression models is employed. The paper ultimately establishes a predictive model capable of predicting the flops and blockbusters with few discrepancies.

\subsubsection{Film Production Clusters
}
Most papers in this section however are concerned with the organization of the film production sector. In an early network analysis, \cite{Kratke_2002} delves into the examination of production clusters within the Potsdam/Babelsberg region of Germany. The primary emphasis lies in utilizing networks as a methodological approach for scrutinizing regional clusters. Through a comprehensive survey, the author identified a total~of 46 companies which comprise the network nodes. The study assesses seven cluster attributes, encompassing functional differentiation, network density, network cohesion, network~centralization, supra-regional transactions, the spatial density of cluster firms, and the quality of the cluster’s institutional infrastructure. The paper concludes that the Potsdam/Babelsberg region aligns with the concept of an “ideal cluster”, and suggests that the network-based methodology provides useful analytical criteria for the evaluation of production clusters. \cite{Xin_2020} also look at networks of production companies in the top-grossing films from Germany (2012-2017), comparing them to company networks in China (2002-2017). Their interest is in the role of formal institutions, including state actors, in the film production process, and they compare the contrasting economic contexts of Germany and China. As well as the production company network, they look at the location of the companies and create an aggregated network of geographic units. They find that formal institutions play a stronger role in the Chinese production network than the German one, in line with their expectations.

\cite{Zhou_2022} also investigates the structure of the Chinese film industry from a network perspective. Using data on the top 20 grossing Chinese films from 2010-2020, the author performs some simple graph analysis of three related networks: the two-mode film director to production company network, the one-mode company-level projection, and the one mode city-city network based on the locations of companies in the company network. They calculate centrality measures and identify cohesive subgroups to understand the structure of these networks, concluding that the networks are characterized by power imbalances, as a “minority of directors and companies dominate the majority of film resources”. They find that these dominant players primarily represent production cultures in the traditionally dominant economic centers of Hong Kong, Beijing and Shanghai. However, based on the spatial patterns in the city network, they suggest that China’s shifting economic landscape has possibly begun to see influence dispersed throughout several emerging eastern cities, dislodging the dyadic control of Beijing and Shanghai.

\cite{Cattani_2008} investigate how network structures of organizations in the film industry might be linked to consensus and legitimacy among production companies. The authors put forth the following hypotheses: (i) greater network connectivity will diminish the exit rates of production companies. This arises from the idea that well-connected networks tend to converge on shared understandings, thanks to reduced gaps in knowledge and information diversity; (ii) repetitive interactions between organizations will likewise reduce the exit rates of production companies. This results from the formation of norms through trial and error, where repeated interactions increase opportunities for these normative processes; and (iii) a higher turnover rate among distribution companies will increase the exit rates of production companies. This stems from the potential for newcomers to disrupt stability and shared understandings. The study tests these hypotheses by constructing a network that represents production companies and the distributors~with which they collaborated on films, covering the period from 1912 to 1970 in Hollywood. The resulting network statistics are then used as independent variables within a survival model for the production companies, ultimately substantiating the proposed hypotheses.

\subsubsection{Co-productions as Networks}
Two papers further examine co-production networks. \cite{Hoyler_2018} investigate urban relationships by applying network analysis to co-productions within~the film industry. They focus on four key markets: China, Germany, Brazil, and France. The study involves collecting data on production companies and their locations from the top 200 highest-grossing films in 2013 and 2014. Using Gephi, they employ community detection and degree centrality to measure node importance. The findings indicate that film production networks are shaped by pre-existing city relationships. France exhibits a monocentric structure, while Brazil and China have a dyadic pattern, and Germany is characterized by a polycentric model. Additionally, established film production hubs such as Los Angeles, London, and Tokyo are consistently present in all analyzed networks. \cite{Zhang_2020} examine the influence of the Chinese “Closer Economic Partnership Arrangement” (CEPA) on co-production networks between Mainland China and Hong Kong. Their simple network analysis revealed that a few studios in both regions held dominance in terms of the number of films produced, degree centrality, and betweenness centrality. They argue that the less centralized filmmaker collaboration network~portrays a scenario in which Mainland China, primarily through producers, wields more control over the financial and administrative aspects of the industry, while Hong Kong, particularly through directors, exerts greater creative and artistic influence.

\subsubsection{Distribution Networks
}
\cite{Rocha_2018} analyze the Brazilian film industry using social network analysis to evaluate the relationships between production and distribution companies. To~construct the network, the authors identified film production and distribution companies with associated films released between 1995 and 2015. They calculate measures of cohesion (density, centralization) and centrality (degree, betweenness, closeness). Despite assessing 613 production companies, only a handful of these companies have cooperative relationships with each other. The lack of cooperative relationships between companies is generally present throughout the results, suggesting that the overall industry~in Brazil is fairly restricted and dispersed. In their 2019 study, \cite{Zhang_2019} also analyze project alliance networks in the film industry using a dataset of production and distribution companies. They sought to understand how alliance network characteristics influenced project performance, specifically examining company-level alliances in both production and distribution. Employing a regression model with box office performance as the dependent variable and network properties as independent variables, the study made several key findings: financial success correlated significantly with alliance network properties, increased degrees of production companies in production alliances positively impacted box office performance, while the degrees of distribution companies in distribution alliances had no significant influence. Interestingly, eigenvector centrality and betweenness centralities in both types of networks did not strongly affect box office performance. Notably, structural holes in distribution alliance networks had a positive impact on box office success, while those in production alliance networks did not. In conclusion, this research highlights the differing effects of network properties in production and distribution alliances on box office performance, with production alliances exerting a stronger influence.

\cite{Choi_2012} are also interested in investigating the intricate system of international film distribution, treating it as a trade network with the objective of shedding light on the flow of film products among developed nations. They discern whether there is a discrepancy between imports and exports in international trade, whether there is a temporal pattern to the networks, and whether there is a geolinguistic clustering among countries in the networks. To test these research questions, they use OECD data on international trade to construct adjacency matrices representing imports and exports of films between countries in two time periods: 1996-2001 and 2002-2006. These networks contain 32 nodes including 30 OECD countries plus China and Hong Kong. They use QAP correlation, eigenvector centrality, multidimensional scaling (MDS) and cluster analysis to analyze their trade networks. They find a moderate to large asymmetry in the import and export networks which is larger in the second time period than the first. In both exports and imports, the US and Canada had the highest centrality scores, and the centrality score ranks were highly correlated between the two time periods, especially for exports. Their MDS reveals close links in all cases between the US and Canada, and among the United Kingdom and France in the import networks, with Asian and Eastern European countries as well as Mexico appearing on the periphery. Their hierarchical clustering analysis (which deliberately omitted the United States) found a group of Commonwealth countries, an East Asian group and smaller European clusters in the early export network, while the later export network was split between a European group (with Canada) and an East Asian group. In a similar vein, \cite{Arrowsmith_2016} use a large global dataset of cinema showtimes to analyze the reciprocity of cultural exchanges by calculating the import and export of films at the national level. The study is particularly interested in reciprocity as a largely overlooked element in understanding cultural exchange and diversity. Informed by digital humanities, network analysis, cultural economics and geo-spatial sciences, the authors highlight those national relationships of exchange that cross-cut traditionally perceived unilateral cultural flows. They find that countries that are close to each other in terms of language, population size and economic development are more likely to have a more equitable and reciprocal relationship in the cinematic exchange network.

Also focusing on distribution networks, \cite{Ehrich_2022} depict the film festival~sector as a network, with festivals connected via film flow (one-mode network) or films linked to festivals (two-mode network). This analysis targets the structural persistence of gender inequality in the sector, using data from 1,523 film festivals and 1,323 films in the 2013 circuit, collected from the “Film Circulation in the International Festival Network and the Influence on Global Film Culture” project. Gender is attributed to a film based on the core creative team’s gender ratio (writer, director, producer), determined computationally using GenderizeR and Gender Guesser. The study views festivals as network agents (nodes) and films as connections, focusing on aggregate social structures rather than individuals. By examining degree distributions among films from women-only, men-only, and mixed-gender core creative teams across festivals and comparing various statistical measures, the study reveals that festivals screening women-led films exhibit less overlap in their film selections compared to those showcasing predominantly male-led films. A select few films from male-led teams circulate more successfully across festivals.

\subsubsection{Reward Networks
}
In a unique study that proposes film awards as a node, \cite{Liu_2022} examine the dynamics of the film industry’s award system to not only explore the growth of the award system itself, but to also understand what leads to a performer’s early career success and initial nominations. They create two networks: an award network (where awards are nodes and the links the number of performers winning both awards), and a performers’ collaboration network. They utilize a logistic regression model to predict a performer’s propensity of prize-winning (the probability that they will win at their first nomination). Their findings show, firstly, that historical performance positively influences an actor’s future award-winning. Furthermore, the longer the performer’s career, the higher likelihood that they will win awards. The second main finding is the performer’s collaborator influence. Actors who had already won awards had a positive effect on a performers’ prize-winning. However, collaborating with directors had a negative effect. The authors conclude that their results indicate that “success breeds success” in the film industry.
\begin{table}[t]
\caption{Feature summary of reported film network analysis with other entities as nodes.}\label{tab1}%
\begin{tabular}{@{}lp{1.45cm}p{2.33cm}p{3.9cm}@{}}
\toprule
Reference & Team Size\footnotemark[1] & Data Size\footnotemark[2] & Evaluation Methodology \\
\midrule
\cite{Kratke_2002} & 1 (M)   & $55$ companies  & - Network analysis   \\
\cite{Cattani_2008} & 4 (M)   & -  & - Regression   \\
\cite{Doshi_2010} & 4 (M) & -  & - Regression \\
\cite{Choi_2012} & 3 (M) & $32$ countries &  - Network analysis, correlation analysis \\
\cite{Arrowsmith_2016}  & 4 (M) & - &  - Network analysis \\
\cite{Rocha_2018}  & 4 (W)  & $613$ companies  & - Centrality analysis \\
\cite{Hoyler_2018} & 2 (M) & -  & - Network analysis \\
\cite{Zhang_2019} & 5 (-) & $86,503$ companies  & - Network analysis, regression \\
\cite{Zhang_2020} & 2 (-) & $243$ studios  & - Network analysis \\
\cite{Xin_2020} & 2 (W) & $460$ firms  & - Network analysis \\
\cite{Ehrich_2022} & 4 (W) & $1,523$ film festivals  & - Network analysis \\
\cite{Liu_2022} & 2 (W) & $3,634$ awards  & - Network analysis, regression \\
\cite{Zhou_2022} & 1 (W) & $47$ cities, $455$ companies  & - Network analysis \\
\botrule
\end{tabular}
\label{table_3}
\end{table}

\section{Discussion and Conclusions}
\label{sec5}
\fontdimen2\font=0.54ex
There are several preliminary conclusions we can draw about the application of network science and theory to film industry analysis and in this context, we also propose avenues for future research. First, our survey reveals that in the initial 25 years of film network scholarship, analysis has focused primarily on networks of individuals and, to a lesser extent, films. In more recent years, a notable diversification has since transpired, and the analytical scope has expanded to include production companies, geographic units, and various hybrid combinations of these entities. As such, film network research is now able to offer a broader and more comprehensive understanding of the complex network dynamics within the film industry. This is promising, but more integration of theory from the relevant domains is needed to push the network analysis beyond simple descriptions of static network structure.

Second, we note that a disproportionate amount of the literature (especially the formative literature for theory-building in this area) is focused on the film industry of the United States, and more work is needed to see whether other film industries are governed by the same network principles. This focus on Hollywood may well be related to the geographic profile of who is contributing to this emerging literature, as a review of the institutional affiliations of authors shows that this literature is dominated by scholars from the Minority World (particularly the United States, Europe and Australia). In this respect, it is encouraging to see that in recent years there has been a growth in the number of contributions to this research area from authors in the Majority World, including analyses of film industries outside of the dominant global export markets. However, we note that the flow of recognition and scholarly engagement is uneven, with authors in the Minority World citing the contributions of scholars in the Majority World far less than they are cited by them.

Third, many of the studies in this survey use historical data and their conclusions can be re-evaluated with more recent data. The film industry and its practices shift over time and should be re-evaluated empirically. Studies can also be examined to test the reproducibility of results. Related to this point, only a few of these studies have made their datasets available. More research in this space could embrace open science principles, especially given that the datasets are typically constructed from public sources with minimal ethical concerns.

 Fourth, our review highlights that, with a few valuable exceptions, very little scholarship has been done which looks at inequalities in the film industry from a relational perspective. Qualitative and non-relational quantitative evidence (not reviewed in this paper) persistently highlights how inequitable these industries and their workforces are. Moreover, the foundational work we reviewed by Faulkner and Jones provides rich evidence of how important reputational and relational mechanisms are for structuring opportunities in  project-based cultural industries. These factors intersect with known network mechanisms of exclusion such as homophily and familiarity to systemically disadvantage some people and benefit others. Despite this, relatively few papers take a network analytic approach to studying film industry inequalities, which in turn makes it harder to develop evidence-based  policies that aim to redress the industry's uneven relationships.

 Fifth, much of the literature relies on one-mode projections of co-affiliation data. Primarily, scholars have analyzed the people-to-people projection of the co-affiliation data, though some papers have analyzed the film-to-film projection as well. Each of these projections offers a limited understanding of the personal networks that organize the industry, or the intertextual structures that emerge from the flow of people between film projects. More methodological sophistication in addressing the inherent multimodality of this type of data is needed. More work is also needed which accounts for the dependencies between different types of nodes (e.g. people and places) that comprise film industry network structure. While some recent contributions have analyzed different types of nodes within the same study, these analyses have largely treated each type of node separately rather than aiming to represent them within a complex multi-level network structure. We hope that scholars will use advances in multi-level network methodology and theory to help model the multi-level nature of actors within the global film industries.

 Finally, more engagement with existing scholarship is needed, especially among researchers from computational disciplines who wish to use the film industry as a case study for advanced methods. There is enough literature that we should expect more cross-citation among the contributions. However, papers continue to appear which apply network analysis to the film industry and yet do not acknowledge any of the earlier work in this area. For the research area to continue to develop, scholars need to properly recognize, engage with and build on the vast amount of scholarship already undertaken. We hope that this survey paper can help simplify this process for scholars wishing to contribute further to this fast-growing topic.

 Together, these findings contribute a synthesis of an important and popular area of network science application that has not yet been systematically reviewed to find coherent themes and gaps. We have highlighted several ways in which applied network scientists can advance understanding of the film industry, and flagged areas that our survey reveals as being in particular need of more attention. There are also several limitations to the present survey. Firstly, the restrictions we placed on our literature search, including the decision to only review publications in English, helped ensure that a survey was feasible and bounded, but it inevitably means that we will not have included some valuable contributions in this space. Furthermore, as the global film industry grapples with technological, business and social challenges we acknowledge the types of nodes we have outlined and their relational dimensions will also evolve and we look forward to the terms of this survey being extended in future iterations of film industry network analysis.

\section*{Declarations}

\begin{itemize}
    \item Funding: This article draws on research supported by the Social Sciences and Humanities Research Council of Canada (SSHRC). 
    \item Conflict of interest/Competing interests: The authors declare that they have no competing interests.
    \item Ethics approval: Not Applicable
    \item Consent to participate: Not Applicable
    \item Consent for publication: All authors read and approved the final manuscript.
    \item Availability of data and materials: Not Applicable
    \item Code availability: Not Applicable
    \item Authors' contributions: Conceptualization: AD, PJ, and DV provide the main idea of the proposed method. Methodology: the methodology and experiments were designed by AD, PJ, and DV. Data collection: literature review was conducted by VV, AK, SH, and AKK and supervised by AD, PJ and DV. Validation: the accuracy of results was checked by AD and PJ. Software: AD implemented the methods and carried out the experiments. Writing-original draft: the original draft was prepared initially by AD, PJ, and DV. Visualization: AD provided all the figures and conceptually checked by PJ and DV. Supervision: the whole project was supervised by DV. Proofread: the paper documents were proofread by all authors.
\end{itemize}


\bigskip






\bibliography{sn-bibliography}

\end{document}